\documentclass{article}
\usepackage{spconf,amsmath,graphicx}
\usepackage[utf8x]{inputenc}
\usepackage{ucs, cite, amsfonts, amssymb, amsthm, bm, bbm, booktabs, algorithmic, algorithm, xcolor, balance,subfigure}
\usepackage[american]{babel}
\usepackage[T1]{fontenc}
\theoremstyle{remark}

\DeclareMathOperator*{\argmax}{\arg\;\max}

\graphicspath{{figures/}}

\title{Subspace-Based Detection in OFDM ISAC Systems under Different Constellations}

\name{Yangming~Lai\textsuperscript{1}, Musa~Furkan~Keskin\textsuperscript{2}, Henk~Wymeersch\textsuperscript{2}, Luca~Venturino\textsuperscript{3}, Wei~Yi\textsuperscript{1}, Lingjiang~Kong\textsuperscript{1}
\thanks{The work of Yangming~Lai, Wei~Yi, and Lingjiang~Kong was supported in part by the scholarship from China Scholarship Council (CSC) under the Grant No.~202206070041, in part by the National Natural Science Foundation of China under Grant 62231008, 62301127 and U19B2017, and in part by the Fundamental Research Funds of Central Universities under Grant ZYGX2020ZB029. The work of Musa~Furkan~Keskin and Henk~Wymeersch was supported by Vinnova Grant 2021-02568. The work of Luca~Venturino was supported by the European Union in the NextGenerationEU plan through the program ``Bando PRIN 2022,'' D.D. 104/2022 (PE7, project ``CIRCE,'' code H53D23000420006).} }
\address{\textsuperscript{1}School of Information and Communication Engineering, University of Electronic Science and\\ Technology of China, Chengdu 611731, China\\
\textsuperscript{2} Department of Electrical Engineering, Chalmers University of Technology, \\ SE 41296 Gothenburg, Sweden\\
\textsuperscript{3} Department of Electrical and Information Engineering, University of Cassino and Southern Lazio,  \\  03043 Cassino, Italy\\}

\begin{document}
%
\maketitle

\begin{abstract}
 
This paper investigates subspace-based target detection in OFDM integrated sensing and communications (ISAC) systems, considering the impact of various constellations. To meet diverse communication demands, different constellation schemes with varying modulation orders (e.g., PSK, QAM) can be employed, which in turn leads to variations in peak sidelobe levels (PSLs) within the radar functionality. These PSL fluctuations pose a significant challenge in the context of multi-target detection, particularly in scenarios where strong sidelobe masking effects manifest. To tackle this challenge, we have devised a subspace-based approach for a step-by-step target detection process, systematically eliminating interference stemming from detected targets. Simulation results corroborate the effectiveness of the proposed method in achieving consistently high target detection performance under a wide range of constellation options in OFDM ISAC systems.

\end{abstract}

\section{Introduction}
\label{sec:intro}

In recent years, the popularity of integrated sensing and communications (ISAC) has surged due to the rapid expansion of spectrally co-existent radars and communication systems in 5G and advanced wireless networks\cite{zheng2019radar,liu2020joint,liu2022integrated}. Orthogonal frequency-division multiplexing (OFDM) waveform, which is an excellent candidate for ISAC transmission\cite{memisoglu2023waveform}, has been widely studied because of its wide application in communication systems and high performance in radar detection\cite{zhang2020joint,keskin2021mimo,keskin2023monostatic}.

Generally speaking, a key step in the communication-functions realization of OFDM waveform is to employ different modulation formats (constellations) such as quadrature amplitude modulation (QAM) or phase shift keying (PSK) to modulate the data onto each subcarrier\cite{cho2010mimo}. The choice of constellations depends on the communication system requirements and signal-to-noise ratio (SNR) considerations.  
However, it is important to note that different constellations can result in varying peak sidelobe levels (PSLs) when implementing radar functions \cite{temiz2020dual,baudais2023doppler}. These variations can bring challenges for target detection, which is a fundamental function of the ISAC system. This is particularly true when there are significant differences in the amplitude of multiple targets, even when using windowing techniques.
To the best of our knowledge, how to achieve high and consistent target detection performance for different constellation options in the OFDM ISAC system has not been investigated.

In this study, we explore subspace-based target detection in OFDM ISAC systems while considering the impact of different constellations. We begin by providing a detailed overview of the OFDM ISAC system, which is capable of fulfilling both radar and communication functions through the use of various constellations. Subsequently, we introduce a subspace-based detection method designed to mitigate sidelobes from strong targets. Finally, we present a series of simulation results that showcase the effectiveness and robustness of our proposed method, particularly in achieving high and consistent target detection performance across various constellation options.

\vspace{-0.2in}
\section{System Description}\label{sect:System and Signal models}
\vspace{-0.1in}
\subsection{OFDM ISAC Signal Model}
We consider a monostatic OFDM ISAC system equipped with one transmitter and one radar receiver, and assume that the radar receiver has perfect knowledge of the transmitted data~\cite{bicua2016generalized,keskin2023monostatic}. In addition, full-duplex operation without self-interference can be realized via sufficient isolation of transmit and receive antennas \cite{Baquero2019TMTT}. The complex baseband signal of an OFDM communication frame with $N$ subcarriers and $M$ symbols can be expressed as
\begin{equation}\label{st}
s(t)=\frac{1}{\sqrt{N}}  \sum_{m=0}^{M-1}\sum_{n=0}^{N-1} h_{n,m} e^{j 2 \pi n \Delta f t  } \Pi\left(\frac{t-mT_{\text{sym}}}{T_{\text{sym}}}\right)
\end{equation}
where $\Pi(t)=\left\{\begin{array}{lr}
1, & t \in[0,1] \\
0, & \text { otherwise }
\end{array}\right.$
and $h_{n,m}$ represents the complex communication data symbol corresponding to the $n$-th subcarrier for the $m$-th symbol, which can belong to different constellations (e.g., PSK, QAM). Besides, $T_{\text{sym}}$ is the total OFDM symbol duration, i.e., $T_{\text{sym}}=T_{\text{cp}}+T$ where $T_{\text{cp}}$ and $T$ denote the durations of the cyclic prefix (CP) and of the OFDM symbol, respectively. The subcarrier spacing and the total bandwidth are $\Delta f=1/T$ and $B=N\Delta f = N/T$, respectively. Then the upconverted transmit signal over the block of $M$ symbols can be represented as $\Re\left\{s(t) e^{j2 \pi f_c t}\right\}$, where $f_{c}$ is the carrier frequency.

We assume that there are $K\geq 1$ point-like targets with round-trip delays $\tau_{k}$, Doppler shifts $\nu_{k}$ and complex amplitudes $\alpha_{k}$ ($k=1, ..., K$) in the surveillance area of the OFDM ISAC system. Then the received continuous-time passband backscattered signal can be expanded as
\begin{equation}
\label{r_t_1}
\Re\left\{  \sum_{k=1}^{K} \alpha_{k} s(t-\tau_{k}(t))e^{j\left[2 \pi f_c (t-\tau_{k}(t))\right]}\right\}
\end{equation}
where $\tau_{k}(t)=\tau_{k}-\nu_{k} t$ is the time-varying delay. Then we can write the baseband signal by downconverting the passband signal in \eqref{r_t_1} and making the narrowband approximation $s(t-\tau_{k}(t)) \approx s(t-\tau_{k})$ as 
\begin{equation}
 \label{r_t_2}
r(t)=\sum_{k=1}^{K}\alpha_{k} s(t-\tau_{k}) e^{-j 2 \pi f_c \tau_{k}} e^{j 2 \pi f_c \nu_{k} t}.   
\end{equation}
Here, we assume that the Doppler-induced phase rotation within an OFDM symbol duration is negligible, i,e, $f_c \nu_k T_{\text {sym }} \ll 1$ so that we have $f_c \nu_k t \approx f_c \nu_k m T_{\text {sym }}$. Then after removing the CP, sampling $r(t)$ at $t=mT_{\text{sym}}+T_{\text{cp}}+\ell T/N$ for $\ell=0, \ldots, N-1, m=0, \ldots, M-1$, and absorbing the constant term $e^{-j 2 \pi f_c \tau_k}$ into the target amplitude $\alpha_{k}$, we can obtain the discrete-time domain signal for the $m$-th symbol as 
\begin{equation}
\begin{aligned}
\label{r_t_discrete}
r_{m}[\ell]=&\sum_{k=1}^{K}\alpha_{k} e^{j 2 \pi f_c m T_{\text{sym}} \nu_{k} } \\ &\times \frac{1}{\sqrt{N}} \sum_{n=0}^{N-1} h_{n,m} e^{j 2 \pi n \frac{\ell}{N}}  e^{-j 2 \pi n \Delta f \tau_{k} }. 
\end{aligned}
\end{equation}
Here, we denote by 
\begin{equation}
\mathbf{b}(\tau_{k}) \triangleq\left[1, e^{-j 2 \pi \Delta f \tau_{k}}, \ldots, e^{-j 2 \pi(N-1) \Delta f \tau_{k}}\right]^\top
\end{equation}
and 
\begin{equation}
\mathbf{c}(\nu_{k}) \triangleq\left[1, e^{-j 2 \pi f_c T_{\text{sym}} \nu_{k}}, \ldots, e^{-j 2 \pi f_c(M-1) T_{\text{sym}} \nu_{k}}\right]^\top
\end{equation}
the frequency-domain and {\color{black}temporal} steering vectors, respectively. Accounting for the existence of noise, the fast-time vector for the $m$-th OFDM symbol is obtained as
\begin{equation}
\begin{aligned}
\label{r_t_discrete_matrix}
\mathbf{r}_{m}=\sum_{k=1}^{K}\alpha_{k}\mathbf{F}_{N}^{H}\big(\mathbf{h}_{m}\odot(\mathbf{b}(\tau_{k})[\mathbf{c}^*(\nu_{k})]_{m} )\big)+\mathbf{z}_{m}\in\mathbb{C}^{N\times 1}
\end{aligned}
\end{equation}
where $\mathbf{F}_{N}\in\mathbb{C}^{N\times N}$ is the unitary DFT matrix with $[\mathbf{F}_{N}]_{\ell,n}=\frac{1}{\sqrt{N}} e^{-j 2 \pi n \frac{\ell}{N}}$,  $\mathbf{z}_{m}\in\mathbb{C}^{N\times 1}$ is the additive noise matrix with $\operatorname{vec}\left(\mathbf{z}_{m}\right) \sim \mathcal{C N}\left(\mathbf{0}, \sigma^2 \mathbf{I}\right)$ and $\mathbf{h}_{m}\triangleq[h_{0,m},\ldots, h_{N-1,m}]^{\top}$.

By aggregating \eqref{r_t_discrete_matrix} across M symbol intervals, the received OFDM ISAC signal over a frame can be expressed as
\begin{equation}
\begin{aligned}
\label{r_t_discrete_matrix_Msymbols}
\mathbf{R}=\sum_{k=1}^{K}\alpha_{k}\mathbf{F}_{N}^{H}\big(\mathbf{H}\odot(\mathbf{b}(\tau_{k})\mathbf{c}^{H}(\nu_{k}) )\big)+\mathbf{Z}\in\mathbb{C}^{N\times M},
\end{aligned}
\end{equation}
where $\mathbf{R}, \mathbf{H}, \mathbf{Z} \in\mathbb{C}^{N\times M}$,  $\mathbf{R}\triangleq[\mathbf{r}_{0},\ldots,\mathbf{r}_{M-1}]$, $\mathbf{H}\triangleq[\mathbf{h}_{0},\ldots,\mathbf{h}_{M-1}]$, and $\mathbf{Z}\triangleq[\mathbf{z}_{0},\ldots,\mathbf{z}_{M-1}]$.

\subsection{Standard FFT-based CFAR Detector}

Here we employ the standard FFT-based method in \cite{braun2014ofdm} to process the matrix $\mathbf{R}$ in \eqref{r_t_discrete_matrix_Msymbols} to obtain range-Doppler data planes. Then we apply the cell average constant
false alarm rate (CA-CFAR) detector in \cite{richards2005fundamentals} to detect the targets. The specific details of the standard FFT-based CFAR detector are omitted here due to space limitations of this paper.


\section{Subspace detection method }\label{sect: MT-JDL}
In this section, we first reorganize the received signal in a vector form and then propose a subspace-based method to mitigate sidelobes from strong targets in OFDM ISAC systems across various constellations.

\subsection{The Reassembly of Received Signals}

The received OFDM ISAC signal in \eqref{r_t_discrete_matrix_Msymbols} can be reassembled as follows
\begin{equation}
\begin{aligned} \label{y}
 \bm{y}&\triangleq\left[\begin{array}{c} \mathbf{r}_{0} \\ \vdots \\ \mathbf{r}_{M-1} \end{array}\right]  \in \mathbb{C}^{N M \times 1} \\
&=\left[\begin{array}{c}\sum_{k=1}^{K} \alpha_k \mathbf{F}_{N}^{H} \mathbf{h}_{0}\odot\mathbf{b}\left(\tau_k\right)\left[\mathbf{c}^*\left(\nu_k\right)\right]_0+\mathbf{z}_{0} \\ \vdots \\ \sum_{k=1}^{K} \alpha_k \mathbf{F}_{N}^{H}\mathbf{h}_{M-1}\odot\mathbf{b}\left(\tau_k\right)\left[\mathbf{c}^*\left(\nu_k\right)\right]_{M-1}+\mathbf{z}_{M-1} \end{array}\right]  \\
 &= \sum_{k=1}^{K} \alpha_k\bm{s}(\bm{x}_k)+\bm{z}
 \end{aligned}
\end{equation}
where $\bm{z}\triangleq[\mathbf{z}_{0}^{\top},\ldots,\mathbf{z}_{M-1}^{\top}]^{\top}\in \mathbb{C}^{N M \times 1}$ is a complex circularly-symmetric Gaussian vector with covariance $C_{\bm{z}}=\sigma^2 \mathbf{I}$, $\bm{x}_k=(\tau_k,\nu_k)$, for $k=1,...,K$, which belong to an assumed inspected delay and Doppler shift points set ${\mathcal{G}}$, and
\begin{equation}
  \bm{s}(\bm{x}_k)\triangleq\left[\begin{array}{c}  \mathbf{F}_{N}^{H}\mathbf{h}_{0}\odot\mathbf{b}\left(\tau_k\right)\left[\mathbf{c}^*\left(\nu_k\right)\right]_0  \\ \vdots \\ \mathbf{F}_{N}^{H} \mathbf{h}_{M-1}\odot\mathbf{b}\left(\tau_k\right)\left[\mathbf{c}^*\left(\nu_k\right)\right]_{M-1}
 \end{array}\right] \in \mathbb{C}^{N M \times 1}.
\end{equation}

\subsection{A Subspace Method for Eliminating the Sidelobes of Strong Targets}

From \eqref{y}, it is evident that $\bm{y}$ results from the superposition of noise and an unspecified quantity of signals originated from $K$ targets. Leveraging the design methodologies of \cite{grossi2020adaptive,grossi2021opportunistic,lai2023joint}, we can extract one target from the received signal at each detection iteration, consider it as an additive interference, and subsequently utilize an estimated interference-plus-noise covariance matrix to remove it.
Therefore, we can write the following sequence of composite binary hypothesis tests as
\begin{equation}
\begin{array}{rll}
\mathcal{H}^{(1)}_1:\;\; & \bm{y}= \bm{s}(\bm{x}^{(1)}) \alpha(\bm{x}^{(1)})+ \bm{z} \\[5pt]
\mathcal{H}^{(1)}_{0}:\;\; & \bm{y}=\bm{z},
\end{array}
\end{equation}
if the iteration index $q=1$ and
\begin{equation}\label{bar_H_k}
\begin{array}{rll}
\mathcal{H}^{(q)}_1:\;\; & \bm{y}= \bm{s}(\bm{x}^{(q)}) \alpha(\bm{x}^{(q)})+\hat{\bm{S}}^{(q-1)}\hat{\bm{\alpha}}^{(q-1)} + \bm{z} \\[5pt]
\mathcal{H}^{(q)}_{0}:\;\; & \bm{y}=\hat{\bm{S}}^{(q-1)}\hat{\bm{\alpha}}^{(q-1)} +\bm{z},
\end{array}
\end{equation}
for $q\geq2$, where $\bm{x}^{(q)}$ and $\alpha(\bm{x}^{(q)})$ are the position and amplitude of the $q$-th target, $\hat{\bm{S}}^{(q-1)}=[\bm{s}(\hat{\bm{x}}^{(1)}),\ldots,\bm{s}(\hat{\bm{x}}^{(q-1)})]^\top$, $\hat{\bm{\alpha}}^{(q-1)}=[\hat{\alpha}(\hat{\bm{x}}^{(1)}),\ldots, \hat{\alpha}(\hat{\bm{x}}^{(q-1)})]^\top$, and $\hat{\bm{x}}^{(1)},...,\hat{\bm{x}}^{(q-1)}$ are the estimated positions, the variables $\{\hat{\alpha}(\hat{\bm{x}}^{(i)})\}_{i=1}^{q-1}$ are modeled to follow a Gaussian distribution with a specific variance $\sigma _{\hat{\alpha} \left( {{{\hat{\bm{x}}}}^{(i)}} \right)}^{2}$. In the following, the interference-plus-noise covariance matrix under $\mathcal{H}^{(q)}_{0}, \forall q$ can be written as
\begin{equation}\label{C_x_q-1}
\bm{C}^{(q)}=\begin{cases}{\bm{C}_{\bm{z}}},\,\,\text{if}\,\,q=1
\\\sum\limits_{i=1}^{q-1}\bm{s}\left( {{{\hat{\bm{x}}}}^{(i)}} \right){\sigma _{\hat{\alpha} \left( {{{\hat{\bm{x}}}}^{(i)}} \right)}^{2}}{{\bm{s}}}^{H}\left( {{{\hat{\bm{x}}}}^{(i)}} \right)+{\bm{C}_{\bm{z}}},\,\,\text{for}\,\,q\geq2.
\end{cases}
\end{equation}
Then, the two negative log-likelihood functions are
\begin{equation}\label{icm_nega_log_H_0_k}
 -\ln {{f}_{0}^{(q)}}\big( {\bm{y}} \big)=
 \ln \Big(\pi ^{NM}\det \bm{C}^{(q)} \Big)+\Big\| \Big(\bm{C}^{(q)}\Big)^{-1/2} {\bm{y}}\Big\|^2
\end{equation}
and
\begin{multline}\label{icm_nega_log_H_1_k}
 -\ln {{f}_{1}^{(q)}}\big( {\bm{y}};\bm{x}^{(q)},\alpha(\bm{x}^{(q)})\big)= 
 \ln \Big(\pi^{NM}\det \bm{C}^{(q)} \Big) \\+\Big\| \Big(\bm{C}^{(q)}\Big)^{-1/2}\Big( {\bm{y}}-\bm{s}(\bm{x}^{(q)}) \alpha(\bm{x}^{(q)})\Big)\Big\|^2
\end{multline}
under $\mathcal{H}_{0}^{(q)}$ and $\mathcal{H}_{1}^{(q)}, \forall q$, respectively. Next, based on the generalized information criterion (GIC) \cite{stoica2004model}, we can obtain the decision rule and detailed expansions as\vspace{-0.1in}
\begin{equation}\label{GLRT-sequence_3}
\argmax_{\bm{x}^{(q)}\in \mathcal{G}}\mathrm{J}^{(q)}(\bm{x}^{(q)})\mathrel{\underset{\mathcal {H}_{0}^{(q)}}{\overset{\mathcal {H}_{1}^{(q)}}{\gtrless}}}{\gamma},
\end{equation}
where 
\begin{align}
    \mathrm{J}^{(q)}(\bm{x}^{(q)})&=\ln \frac{{{f}_{1}^{(q)}}\big( {\bm{y}};\bm{x}^{(q)},\alpha(\bm{x}^{(q)})\big)}{{{f}_{0}^{(q)}}\big( {\bm{y}} \big)}\notag\\
    &=\left\|\bm{T}^{(q)}(\bm{x}^{(q)}){(\bm{C}^{(q)})^{-1/2}}\bm{y}\right\|^2 \notag\\
&=\left\|{(\bm{C}^{(q)})^{-1/2}}\bm{s}(\bm{x}^{(q)})\hat{\alpha}(\bm{x}^{(q)})\right\|^2, 
\end{align}\vspace{-0.2in}
\begin{align} \label{project-2}
\bm{T}^{(q)}(\bm{x}^{(q)})=&{(\bm{C}^{(q)})^{-1/2}}\bm{s}(\bm{x}^{(q)})\Bigg\{\bm{s}^{H}(\bm{x}^{(q)}){(\bm{C}^{(q)})^{-1}}\bm{s}(\bm{x}^{(q)}) \Bigg\}^{-1}\notag\\&  \bm{s}^{H}(\bm{x}^{(q)}) {(\bm{C}^{(q)})^{-1/2}},
\end{align} \vspace{-0.2in}
\begin{multline}\label{eq:alpha_MSD_ICM}
\hat{\alpha}(\bm{x}^{(q)})=\left(\bm{s}^{H}(\bm{x}^{(q)}){(\bm{C}^{(q)})^{-1}}\bm{s}(\bm{x}^{(q)})\right)^{-1}\\\bm{s}^{H}(\bm{x}^{(q)}){(\bm{C}^{(q)})^{-1}}\bm{y}
\end{multline}
is the maximum likelihood (ML) estimate of $\alpha(\bm{x}^{(q)})$, and the detection threshold $\gamma$ can be set to satisfy a given probability of false alarm $\text{P}_{\text{fa}}$.

If a threshold crossing occurs, a target detection is declared and the ML estimate of its position, say $\hat{\bm{x}}^{(q)}$, is recovered from the location of the maximum in \eqref{GLRT-sequence_3}. Simultaneously, the ML estimate of the corresponding gain vector is determined as $\hat{\alpha}(\hat{\bm{x}}^{(q)})$. The decision logic executes the test in~\eqref{GLRT-sequence_3} for $q=1,2,\ldots$ until no additional target is found. Finally, we replace the unknown variance $\sigma _{\hat{\alpha} \left( {{{\hat{\bm{x}}}}^{(i)}} \right)}^{2}$ in \eqref{C_x_q-1} by $\sigma _{\hat{\alpha} \left( {{{\hat{\bm{x}}}}^{(i)}} \right)}^{2}={\left\| \hat{\alpha} ( {{{\hat{\bm{x}}}}^{(i)}} ) \right\|}^{2}$. We note that while the proposed algorithm shares similar principles to the SAGE algorithm \cite{SAGE_99}, they differ in the use of covariance matrices for echo elimination and interference cancellation (IC) (i.e., SAGE applies IC only for initialization unlike the proposed algorithm, which performs IC at every iteration).

\section{ Performance Analysis}\label{sect:Simulation example}

\begin{figure}[t] 
		\centering
		\subfigure[Range slice data after power normalization of the first iteration of the proposed method (identical to those of the standard FFT-based method).]{
			\begin{minipage}[h]{0.9\columnwidth}\label{fig: the range dimensional data in a single realization (a)}
	\includegraphics[width=1\textwidth,draft=false]{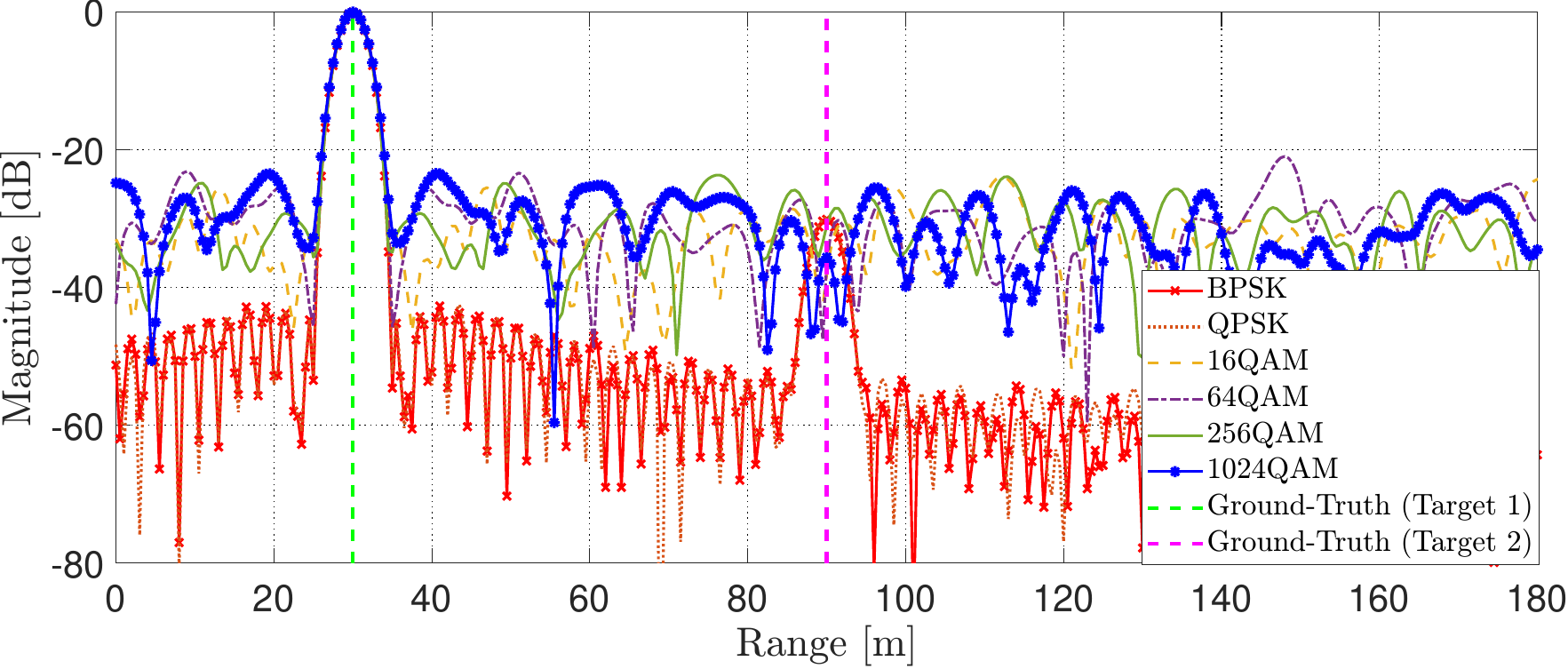}
			\end{minipage}
        }\\
		\subfigure[Range slice data after power normalization of the second iteration  of the proposed method.]{
			\begin{minipage}[h]{0.9\columnwidth}\label{fig: the range dimensional data in a single realization (b)}
	  \includegraphics[width=1\textwidth,draft=false]{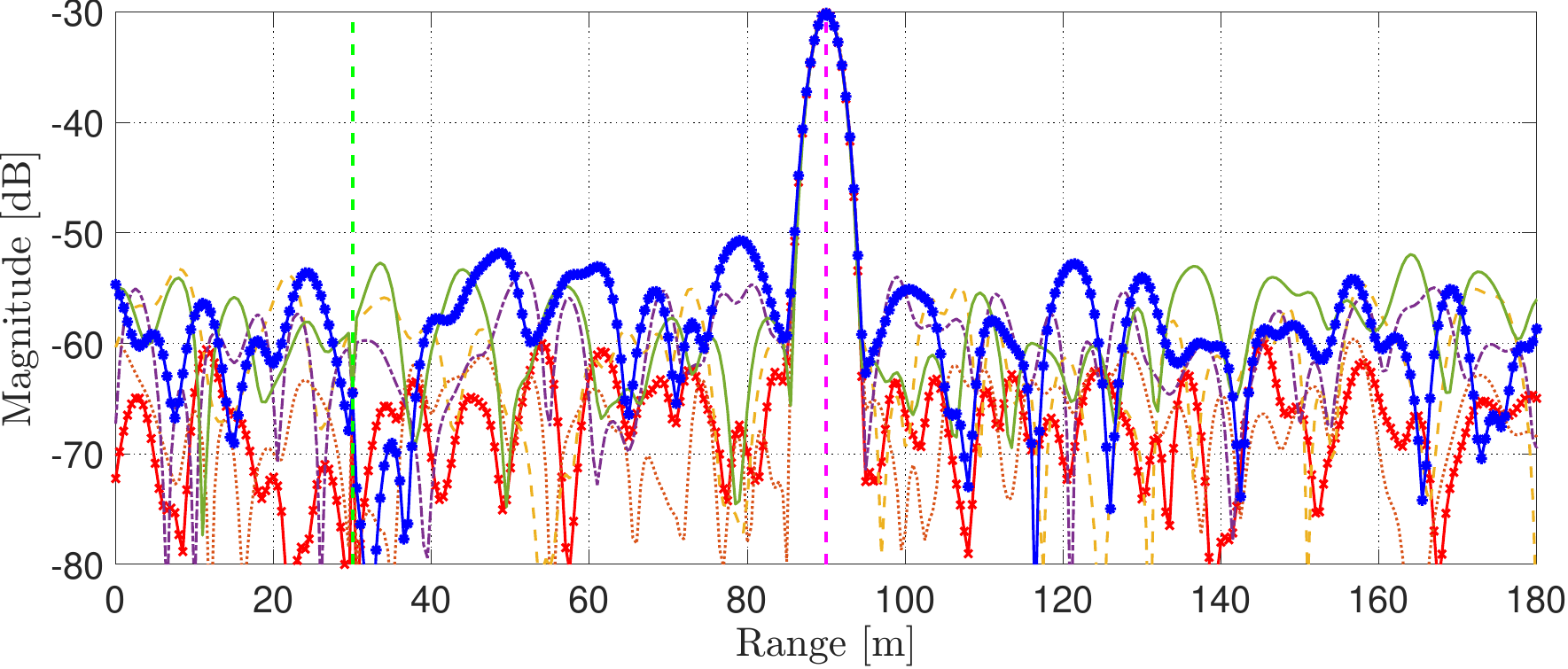}
			\end{minipage}
		}	
	\caption{The range slice of the proposed subspace method and the standard FFT-based method for different constellations.}
		\label{fig: the range dimensional data in a single realization}
  \vspace{-0.2in}
\end{figure}

\begin{figure}[t]
	\centering
	\includegraphics[width=0.75\columnwidth]{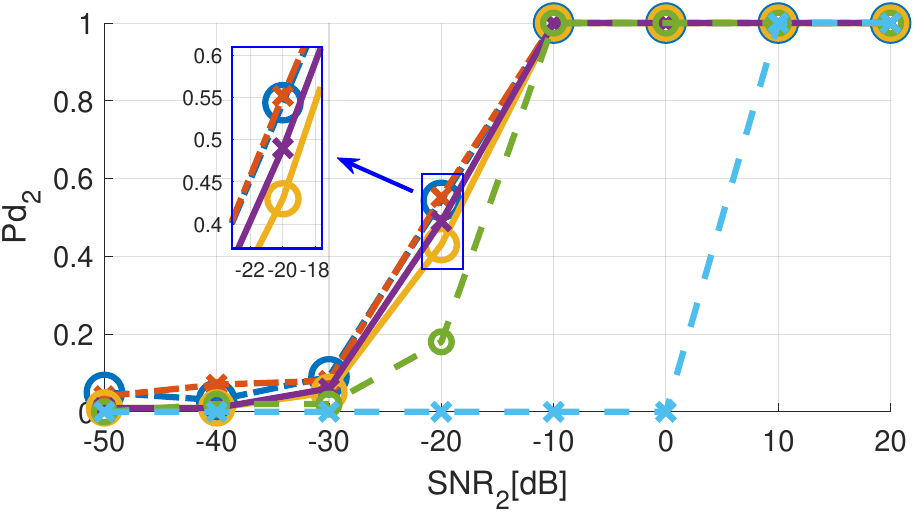}\\
	\includegraphics[width=0.75\columnwidth]{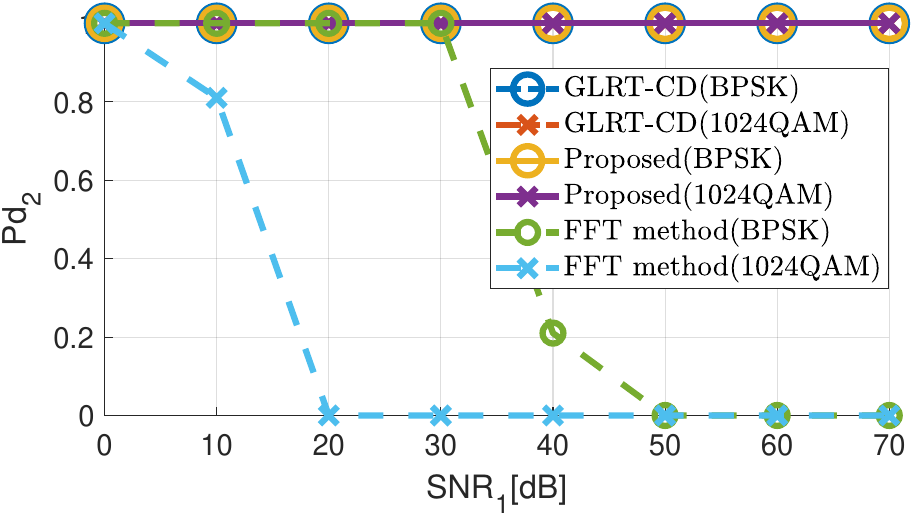}\\
        \includegraphics[width=0.75\columnwidth]{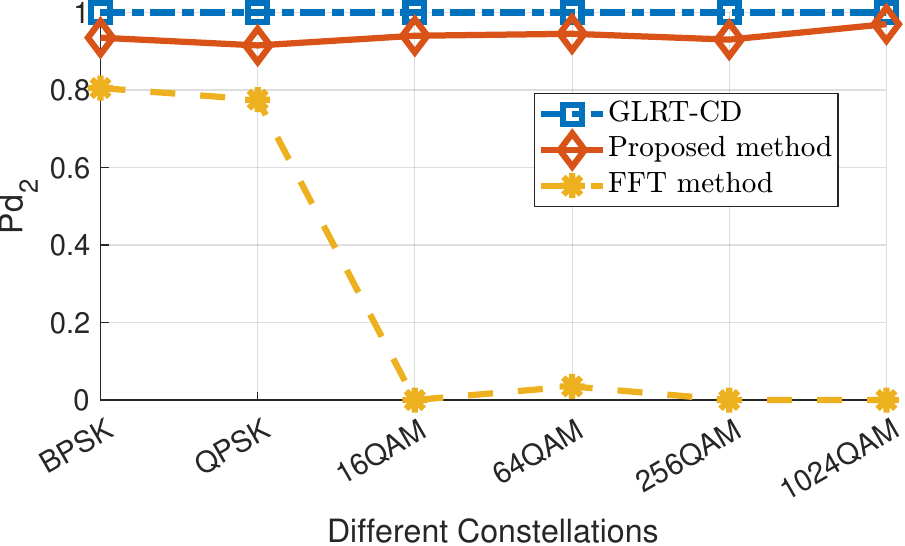}
	\caption{ $\text{P}_{\text{d}}$ of Target 2 versus $\text{SNR}_{2}$ when $\text{SNR}_{1}=30~\text{dB}$ (top), versus $\text{SNR}_{1}$ when $\text{SNR}_{2}=-10~\text{dB}$ (middle), and versus different constellations when $\text{SNR}_{1}=30~\text{dB}$ and $\text{SNR}_{2}=-15~\text{dB}$ (bottom).}
	\label{fig:Fig_monte_carlo_fix_SNR_strong_target}
 \vspace{-0.2in}
\end{figure}

In this simulation study, we consider an OFDM ISAC system with $f_c=28~\mathrm{GHz}$, $B=61.44~\mathrm{MHz}$, $\Delta f=120~\mathrm{kHz}$, $N={512}$, $M={1}$, $T= 8.333~\mu\mathrm{s}$, $T_{\mathrm{cp}}=1.666~\mathrm{\mu s}$, $T_{\text{sym}}=10~\mathrm{\mu s}$, and $\text{P}_{\text{fa}}=10^{-4}$ \cite{Puc22}. Two targets are placed at $30~\mathrm{m}$ (Target 1) and $90~\mathrm{m}$ (Target 2) in front of the system. The considered constellations include the BPSK, QPSK, 16QAM, 64QAM, 256QAM, and 1024QAM. We assume that Target 2 is the one of interest, the system performance is assessed in terms of its probability of detection ($\text{P}_{\text{d}}$), $100/\text{P}_{\text{fa}}$ Monte Carlo (MC) trails are used to set the detection thresholds, and the SNR of the $k$-th target is defined as $\text{SNR}_{k}=|\alpha_{k}|^2/\sigma^2$. 
For comparison, we also include the performance obtained with the standard FFT-based CFAR detector and with the generalized likelihood ratio test with cleaned data (GLRT-CD)\cite{grossi2021opportunistic}, which employs the proposed subspace method when the echoes produced by other targets are ideally removed, also called the single-target benchmark.

We first assume the SNRs of Targets 1 and 2 are {\color{black}40} and 10 $\text{dB}$, and Fig.\,\ref{fig: the range dimensional data in a single realization} shows the range slice comparison of the proposed subspace method and the standard FFT-based method employing the considered different constellation options in a single realization. It can be seen from Fig.\,\ref{fig: the range dimensional data in a single realization (a)} that the main lobes of the weak target (Target 2) are not apparent {\color{black}for QAM constellations} due to the strong sidelobes of Target 1; obviously, it is not easy to detect Target 2 well by employing CFAR detector on this range slice for each {\color{black}QAM} constellation. However, Fig. \ref{fig: the range dimensional data in a single realization (b)} shows that when employing the proposed subspace method, the sidelobes of the detected Target 1 will be removed and the main lobes of Target 2 will be apparent at the second iteration for all constellation options, thus the improvement in its detection performance is expected.

{\color{black}Then, we select the BPSK and {\color{black}1024QAM} constellations to investigate the $\text{P}_{\text{d}}$ of Target 2 versus $\text{SNR}_{2}$ when $\text{SNR}_{1}=30~\text{dB}$ and that versus $\text{SNR}_{1}$ when $\text{SNR}_{2}=-10~\text{dB}$ by MC tests. It can be seen in the top and middle figures of Fig.\,\ref{fig:Fig_monte_carlo_fix_SNR_strong_target} that the standard FFT-based CFAR detector performs more badly, especially when the gap between the strong and weak targets becomes larger. However, the proposed subspace method always performs close to the GLRT-CD, showing its effectiveness and robustness in weak target detection under high sidelobes masking. Finally, the bottom figure of Fig.\,\ref{fig:Fig_monte_carlo_fix_SNR_strong_target} shows the $\text{P}_{\text{d}}$ of Target 2 versus different constellations when $\text{SNR}_{1}=30~\text{dB}$ and $\text{SNR}_{2}=-15~\text{dB}$, we can see the proposed method achieves high and consistent target detection performance for different constellation options.}

\section{Conclusions}\label{sect: Conclusion}

In this paper, we consider the subspace-based target detection in OFDM ISAC system under different constellations. We provided a detailed OFDM ISAC system description and derived a subspace-based procedure to gradually detect targets, upon eliminating the interference caused by the detected targets. Simulation results verified the effectiveness and robustness of the proposed method. Compared to the standard FFT-based CFAR detector, the proposed method performs close to the single-target benchmark and achieve high and consistent target detection performance for different constellation options. Next, we will explore the application of the proposed subspace-based method in more intricate modulation schemes of the MIMO-OFDM ISAC systems. 


\clearpage
\vfill\pagebreak

\bibliographystyle{IEEEbib}
\bibliography{strings,references}
\end{document}